\documentclass{article}
\pdfoutput=1
\usepackage[utf8]{inputenc}
\usepackage{jheppub}

\usepackage{amsmath,amsthm,amsfonts,amssymb}
\usepackage{multirow}
\usepackage{longtable}
\usepackage{mathtools}
\usepackage{comment}
\usepackage{xcolor}

\newcommand{\be}{\begin{equation}}
\newcommand{\ee}{\end{equation}}

\definecolor{Vcol}{rgb}{0.7,0.2,0.9}
\definecolor{Mcol}{rgb}{0.2,0.8,0.6}
\definecolor{macrocol}{rgb}{0.6,0.4,0.4}
\definecolor{macromcol}{rgb}{0.2,0.4,0.8}
\definecolor{TBCcol}{rgb}{0.9,0.3,0.9}
\definecolor{MRcol}{rgb}{0.82, 0.1, 0.26}

\title{Snowmass white paper: Quantum information in quantum field theory and quantum gravity}

\author[1]{Thomas Faulkner,}
\author[2]{Thomas Hartman,}
\author[3]{Matthew Headrick,}
\author[4]{Mukund Rangamani,}
\author[3]{and Brian Swingle}
\date{March 2022}

\affiliation[1]{Department of Physics, University of Illinois, Urbana IL, USA}
\affiliation[2]{Department of Physics, Cornell University, Ithaca NY, USA}
\affiliation[3]{Martin Fisher School of Physics, Brandeis University, Waltham MA, USA}
\affiliation[4]{Center for Quantum Mathematics and Physics (QMAP)\\ 
Department of Physics \& Astronomy, University of California, Davis CA, USA}

\emailAdd{tomf@illinois.edu}
\emailAdd{hartman@cornell.edu}
\emailAdd{headrick@brandeis.edu}
\emailAdd{mukund@physics.ucdavis.edu}
\emailAdd{bswingle@brandeis.edu}

\abstract{
We present a summary of recent progress and remaining challenges in applying the methods and ideas of quantum information theory to the study of quantum field theory and quantum gravity. Important topics and themes include: entanglement entropy in QFTs and what it reveals about RG flows, symmetries, and phases; scrambling, information spreading, and chaos; state preparation and complexity; classical and quantum simulation of QFTs; and the role of information in holographic dualities. We also highlight the ways in which quantum information science benefits from the synergy between the fields.
}

\preprint{BRX-TH-6703}

\begin{document}

\maketitle

\setcounter{section}{-1}

\section{Executive summary}

Quantum information provides a powerful new perspective on the framework of quantum field theory which is agnostic about energy scale, field content, duality frame, and so on, and therefore cuts through the space of physical phenomena in a fundamentally different way from traditional quantities such as correlation functions and scattering amplitudes. Concepts such as entanglement and complexity yield valuable new insights into many aspects of quantum field theories, including correlations, symmetries, RG flows, phases, transport, and thermalization. Moreover, though it is often said that our theory of spacetime and gravity is in tension with quantum theory, recent developments suggest that spacetime and gravity actually emerge from complex patterns of quantum information. This new quantum information perspective also brings with it new approaches to classical simulation, the novel possibility of quantum simulation, and many connections to many-body physics and beyond. 
Conversely, quantum information science also continues to benefit from the synergy between the two fields, with numerous new conceptual insights and calculational tools arising out of questions framed in the study of quantum field theory and quantum gravity.

\section{Introduction and themes}

Quantum field theory (QFT) is the lingua franca of the high-energy world, and is anchored on  principles of quantum mechanics, locality, and relativistic invariance. It is the canonical framework to explain the fundamental laws of nature and has led to the standard models of particle physics and inflationary cosmology. In its (non-relativistic) many-body avatar, it plays a central role in understanding myriad phases of matter and their properties. 

However, QFT, as formulated in its standard usage, does not leverage the full gamut of quantumness inherent in quantum mechanics. Traditionally, one is focused on observables that can be computed using correlation functions of local operators, which can be translated thence into physical quantities such as scattering cross-sections and dynamical response functions. The Wilsonian effective field theory paradigm is also primarily geared towards identifying relevant operators and understanding their correlation functions below some cut-off scale. Furthermore, techniques such as path integrals, the renormalization group, effective theories, symmetries, and dualities are all primarily geared in textbook treatments towards computing such correlation functions.

There is, however, more information to be mined by generalizing the framework to one more cognizant of the underlying quantum mechanical structure. Such a perspective, in particular, has to account for the fact that composite quantum mechanical systems exist in tensor product superpositions, which leads to the essential concept of entanglement. Focusing on these aspects can help us to better understand the field theory framework, and it appears more suited to addressing questions in the quantum gravitational setting. This quantum information perspective on QFT, a subject that has seen remarkable progress in the past decade and a half, is the central focus of our white paper. Much of the progress is thanks to connections to classical and quantum gravity, another theme of this white paper.

A main advantage of applying the quantum information lens to QFT is that it eschews specific field content and observables, and studies rather the information content of the QFT wavefunctionals. Correlation functions are replaced by entropic quantities that quantify total correlations, of all fields, in different spatial regions. This latter statement exploits the inherent locality of interactions in QFT. The resulting description is (in principle) manifestly invariant under field redefinitions and choice of duality frame, and it is suitable for studying a large array of phenomena in QFT, from static to dynamic and weak to strong coupling. 

The general themes that have been explored in this broad area include refinement of conventional field-theoretic tools (e.g.\ path integral methods) to quantify spatially-ordered entanglement in QFTs, quantifying measures of entanglement in mixed states, and the revival of operator algebraic formulations of QFT. In a parallel development, the geometrization of information theoretic measures in the context of the holographic AdS/CFT correspondence has played an important role in furthering our understanding of the holographic dictionary. 
Additionally, progress has been made on questions relating to the complexity of state preparation, which is important for quantum simulations, and it has furthermore been argued that these ideas have a physics role to play in QFT and quantum gravity. 
Many of these developments have been accompanied by novel insights on the quantum information side, showing that the synergy between the fields benefits both sides.

In this white paper, we will give a broad-brush overview of what we view as the most important recent developments and open problems in these areas. In \S\ref{sec:defentqi}, we outline some of the basic definitions, properties, and methods for quantifying quantum information. In \S\ref{sec:symtop} we describe how these ideas have played a role in understanding the structure of the vacuum and the renormalization group in QFTs. In \S\ref{sec:dynamics} we turn to the implications of quantum information for QFT dynamics. In \S\ref{sec:simulation} we outline the prospects for simulating QFT dynamics using both classical and quantum methods. In each of these sections, we emphasize, in italics, what, we believe, are some of the most important open questions in the subject. In \S\ref{sec:outlook}, we conclude with an overview of the major themes that we believe will drive progress over the next decade or so in better comprehending the nature of quantum information in QFT and gravity.

Of course, in such a short review of such a broad and rapidly developing field, we cannot be comprehensive. Among many topics left out are those touching on the black hole information problem, as these are addressed in another white paper \cite{Bousso:2022ntt}. Other white papers on related topics include \cite{Giddings:2022jda,Meurice:2022xbk,Blake:2022uyo}. Some useful pedagogical resources on the subject include \cite{Nielsen:2010aa,Wilde:2013aa} for an introduction to quantum information in quantum mechanics, \cite{Hagg:2012alg,Ohya:2004qnt} for algebraic treatment of QFTs, \cite{Calabrese:2009qy} for computation of entanglement in conformally invariant QFTs, and 
\cite{Nishioka:2009un,VanRaamsdonk:2016exw,Rangamani:2016dms,Headrick:2019eth} for reviews of developments in the context of holography.

\section{Defining and characterizing quantum information}
\label{sec:defentqi}

Much progress has been made in the last 10 years on the problem of defining and characterizing entanglement in QFT and gravity. One central goal is to quantify the amount of entanglement and extract universal contributions that are independent of the chosen regulator. Another central goal is to understand the organization of entanglement in the state, in terms of how it is structured in space and between scales, and how it can be prepared. In this section, we will summarize some of the approaches to these problems.

\subsection{Entanglement entropy: definition and basic properties}
\label{sec:EEdefinition}    

Consider a  QFT on $d$-dimensional Minkowski spacetime. Its Hilbert space is defined on a Cauchy slice $\Sigma$, which can be divided into two disjoint subregions, $\Sigma = A \sqcup A^c$. The Hilbert space on $\Sigma$ is  assumed, for now, to factorize $\mathcal{H}_\Sigma = \mathcal{H}_A \otimes \mathcal{H}_{A^c}$. This factorization requires a UV regulator with length scale $\epsilon$. A pure-state wavefunction on $\Sigma$, reduced to the region $A$, is described by a local density matrix $\rho_A$ that captures all correlation functions inside $A$. Such a density matrix is necessarily mixed, indicating entanglement between the two regions.  Its von Neumann entropy, the so-called entanglement entropy, quantifies this entanglement:
\begin{equation}
\label{def:ee}
S(A)_\rho := S_{\rm vN}(\rho_A) = - {\rm Tr} \left(\rho_A \ln \rho_A\right) = \frac{\text{Area}(\partial A)}{\epsilon^{d-2}} + \cdots \,.
\end{equation}
Such a setup was first considered in \cite{Sorkin:1984kjy,Bombelli:1986rw} in the context of black hole physics and then revisited in \cite{Srednicki:1993im,Holzhey:1994we}.

The entanglement entropy is UV divergent; it is regulated by a short-distance cutoff $\epsilon$, with the leading divergent term proportional to the area of the boundary of the region $A$, as indicated by the last equality. The boundary of $A$ is called the entangling surface (or entanglement cut). This divergence is a universal consequence of the short-distance correlations present in any QFT \cite{Witten:2018zxz}. In two-dimensional conformal field theories (CFTs), the leading divergence becomes a log, and one finds a particularly universal form for the vacuum entropy \cite{Holzhey:1994we,Calabrese:2004eu}:
\begin{equation}
\label{def:ee2}
S(A)_\rho = \frac{c}{3} \log \frac{L}{\epsilon}\,,
\end{equation}
for a region $A$ of length $L$, with $c$ the central charge of the CFT.

Over the last decade and a half, the structure of such divergent terms in QFT entanglement has been intensively studied; see for example \cite{Nishioka:2018khk}. 
Our current understanding comes from a variety of methods, including free-theory computations \cite{Casini:2009sr}, path integral and heat kernel methods \cite{Solodukhin:2008dh,Hertzberg:2010uv}, CFT methods \cite{Rosenhaus:2014woa}, and AdS/CFT \cite{Solodukhin:2008dh,Myers:2010tj,Liu:2012eea}.
In addition to the leading area divergence, sub-leading terms can contain logs similar to \eqref{def:ee2}. Logarithmic terms appear in even dimensional CFTs for smooth regions and are related to the conformal anomaly of the CFT \cite{Calabrese:2004eu,Solodukhin:2008dh,Safdi:2012sn}. In odd dimensions, logs can appear due to corners or other singularities in the shape of the region $A$ \cite{Casini:2006hu,Casini:2008as,Hirata:2006jx,Klebanov:2012yf}. While divergent area terms are typically dependent on the cutoff procedure, the logarithmic terms are universal and often related to vacuum correlation functions of the CFT. By connecting information quantities to basic data in QFT, we might expect to be able to put new constraints on the QFT theory space, a theme of many of the exciting results in this area discussed below.

Much of the more recent progress in studying entanglement entropy in QFT has been in studying the terms hidden in the ellipses of \eqref{def:ee}. These terms can be finite, state-dependent \cite{Casini:2008cr}, scale-dependent \cite{Liu:2012eea,Faulkner:2014jva,Casini:2014yca}, and shape-dependent \cite{Mezei:2014zla,Faulkner:2015csl,Balakrishnan:2016ttg,Bianchi:2016xvf,Dong:2016wcf}. By considering differences of quantities with the same UV divergence, we can extract these terms. An example is the mutual information:
\begin{equation}
\label{MI:mono}
I(A:B) := S(A) + S(B) - S(AB) \geq 0\,,
\end{equation}
where $A$, $B$ are disjoint regions on the same Cauchy slice and $AB:=A\cup B$. As long as $A,B$ do not share a common boundary, the mutual information is UV-finite. It quantifies the total amount of correlation between regions, including both classical correlations and quantum entanglement. $I(A:B)$ can be used to extract  universal data in CFTs, such as the OPE coefficients and operator dimensions \cite{Headrick:2010zt,Calabrese:2010he,Cardy:2013nua,Agon:2015ftl,Casini:2021raa}, thus providing another connection to more standard QFT data. Furthermore, it satisfies many nice properties, including non-negativity (a consequence of subadditivity of entropy) and monotonicity under deformations of either region (a consequence of strong subadditivity, SSA):
\begin{equation}\label{SSA}
I(A:B) \leq I(A:BC)
\end{equation}
Such quantum information bounds have far-reaching implications for QFT. For example, SSA plays a key role in the proofs of the entropic C-theorems, as we will review in \S\ref{sec:RG}. Another divergence-free entropic quantity is the tripartite information:
\begin{equation} \label{I3def}
I_3(A:B:C) := I(A:B) + I(A:C) - I(A:BC)
\end{equation}
which can be used to extract universal data in a topological QFT from the ground state wavefunction with  a judicious choice of regions $A$, $B$, $C$ \cite{Kitaev:2005dm,LevinWen}.

Another important quantity, the relative entropy, likewise removes the divergence in the entanglement entropy by considering two different states for a single region:
\begin{equation}\label{relentdef}
S(\rho_A | \sigma_A) :=  S(A)_\sigma - S(A)_\rho + \left< K_\sigma \right>_\rho -  \left< K_\sigma \right>_\sigma
\end{equation}
where $K_\sigma := - \log \sigma_A$ is called the modular Hamiltonian. The relative entropy is a generalization of the free energy difference in statistical mechanics, for which $\sigma$ is some thermal density matrix and $K_\sigma = \beta H$ with $H$ the actual Hamiltonian. The relative entropy is a measure of the difference between two states; 
it is non-negative and vanishes if and only if the states are the same.  Like the mutual information, the relative entropy has many nice properties, such as inclusion monotonicity, and as such its study in QFT has many important applications \cite{Casini:2008cr,Blanco:2013joa}. 
 
The universal divergence in \eqref{def:ee} signals a fundamental issue in the study of quantum information in QFT: it is not always possible to factorize the Hilbert space across some geometric cut, as was naively suggested above. There are several approaches to address this. One approach is to consider UV regularizations of the theory, such as a lattice regulator (with the UV cutoff $\epsilon$ being the lattice spacing), for which a factorization can be made manifest \cite{Peschel_2003}. However, for gauge theories, the situation is more complicated. Even with a lattice regulator, the projection to the gauge-invariant Hilbert space does not allow for a local tensor factorization \cite{Buividovich:2008gq,Donnelly:2011hn,Donnelly:2014gva,Casini:2013rba,Ghosh:2015iwa,Lin:2018bud}.\footnote{For various approaches to entanglement in gauge theories, with or without a lattice regulator, see \cite{Kabat:1995eq,Eling:2013aqa,Donnelly:2012st,Agon:2013iva}.} Instead, one can choose to work in an extended Hilbert space \cite{Buividovich:2008gq,Donnelly:2011hn,Ghosh:2015iwa}, where charged degrees of freedom are included at the entangling surface. Such degrees of freedom can be understood as edge modes in topological gauge theories \cite{Fliss:2017wop}. Alternately, one can use an algebraic description at the level of the lattice theory, which relies on local algebras that necessarily have non-trivial centers \cite{Casini:2013rba}. There are ambiguities in the choice of such localized algebras. Entropies can be defined and computed in both cases. While different approaches lead to similar physical conclusions, there have been notable issues with reproducing the same log divergences \cite{Eling:2013aqa,Huang:2014pfa} (related to the $a$-type anomaly coefficient.) Proposed solutions can be found in \cite{Donnelly:2015hxa,Casini:2019nmu}.  

Some QFTs, such as those with chiral fermions or with gravitational anomalies, may not admit a lattice regularization in the first place \cite{Hellerman:2021fla}. Nevertheless, there is progress in studying entanglement entropy in these theories \cite{Wall:2011kb,Castro:2014tta,Iqbal:2015vka,Longo:2019pjj}.

Path integrals also provide a natural UV regulator, by smoothing out the singular spaces \cite{Solodukhin:2011gn, Lewkowycz:2013nqa,Ohmori:2014eia,Anegawa:2021osi} that are used to compute entanglement (see \S\ref{sec:methods}).  This often comes at the expense of a direct connection between the regularized quantity and a von Neumann entropy. 

In the end, one might get the impression that there is too much ambiguity in the study of entanglement entropy in QFT. How can one extract meaningful results?  In fact, it is likely that we should not ascribe too much importance  to the ambiguities in defining entanglement entropy at the UV cutoff or lattice scale. Most of the important results on entanglement entropy in any case derive from the UV finite quantities that were mentioned above. In general, as long as one computes quantities that are UV-insensitive, the ambiguities and disparate approaches mentioned above lead to the same  answers. In particular, due to properties such as monotonicity \eqref{SSA}, the continuum limit is often smooth for quantities like mutual information, as any fluctuations from the lattice scale are absent for monotonic quantities \cite{Casini:2013rba}. Thus, approaches based on quantities that are UV finite from the outset such as mutual information \cite{Casini:2015woa} and reflected entropy \cite{Dutta:2019gen} are the most promising. In fact, these quantities can be given a definition directly in the continuum. We discuss this now.

The (Haag-Kastler) axiomatic framework of algebraic QFT \cite{Haag:1963dh,haag2012local} provides another resolution to the factorization issue directly in the continuum limit. In this case, issues of UV sensitivity are removed from the beginning \cite{Witten:2018zxz}. Moreover, in this approach, it is clear which quantities have a continuum limit in the first place.
The basic idea here is to replace the tensor factorization of the Hilbert space with the algebras of operators associated to the region: $A \rightarrow \mathcal{A}_A$ and $B \rightarrow \mathcal{A}_B$. In finite dimension, one deals with the algebra of all operators on a Hilbert space;  these are examples of type I von Neumann algebras. However, in the continuum, a richer class of operator algebras arises, and these make it impossible to describe the local physics by type I von Neumann algebras. The appropriate algebras are type III$_1$ von Neumann algebras \cite{Witten:2018zxz}, as can be established by using basic properties of short-distance correlations near $\partial A$. Such algebras do not admit pure states localized inside $A$, or a factorized description. So, for example, all states have infinite von Neumann entropy as signaled by the divergence in \eqref{def:ee}. 

In this approach, the density matrix for a local region does not exist from the outset, so we need a new way to characterize quantum information. The new method uses Tomita-Takesaki modular theory. (For a review see \cite{Witten:2018zxz} and for other discussions of modular theory in QFT see for example \cite{Casini:2009vk,Lashkari:2018nsl,Lashkari:2018oke,Ceyhan:2018zfg,Lashkari:2019ixo,Fries:2019ozf}.) For instance, while  
$\rho_A$ does not exist in the continuum,  
$\rho_A \otimes \rho_B^{-1}$ does exist, and is called the modular operator $\Delta$. A definition can be given for $\Delta$, directly from the algebra and the state of interest. Information quantities, such as relative entropy, can now be defined using the modular operator and certain generalizations involving two states \cite{Araki:1976zv}. The entanglement entropy itself, however, fails to exist. The modular operator generates an important automorphism of the algebra called modular flow: $\mathcal{O} \rightarrow \Delta^{is} \mathcal{O} \Delta^{-is}$, 
with $s$ the modular ``time''. If the original state is thermal, then modular flow is simply time evolution. For more general states, it is a powerful state-dependent evolution in an emergent time direction. While many of these tools have been around for half a century, applications to the study of quantum information issues have been limited and are only now being developed more thoroughly \cite{Ciolli:2019mjo, Casini:2019qst,Longo:2019pjj,Longo:2017mbg}. 
There is an interesting connection between analyticity in modular time and causality in the QFT \cite{Balakrishnan:2017bjg} as well as in a putative holographic dual \cite{Faulkner:2018faa}. Rigorous results for modular flow are available for free theories \cite{Longo:2017mbg,Hollands:2019hje}.
 
The mutual information similarly has an algebraic continuum definition. It  relies on a notion called the split property \cite{haag2012local}, whereby if two regions $A$, $B$ do not share a common boundary (in other words there is a third region $C$ that lies between $A$ and $B$ on $\Sigma$), then the algebra associated to $AB$ behaves like a tensor product $\mathcal{A}_A \otimes \mathcal{A}_B$. Quantum field theories with standard thermodynamic behavior are expected to satisfy the split property. This property helps us define another continuum measure, the reflected entropy $S_R$ \cite{Dutta:2019gen}. Reflected entropy was first considered in the context of AdS/CFT, where it was shown to be dual to twice the entanglement wedge cross-section (see \S\ref{sec:mixedstate}). Since then, it has been computed using a variety of methods \cite{Bueno:2020vnx,Siva:2021cgo,Liu:2021ctk,Akers:2021pvd}. Rigorous results computing mutual information and reflected entropy are now available for free theories \cite{Longo:2017mbg,Longo:2019pjj}. 

Finally, quantifying entanglement can be thought of operationally, for example via the number of distillable EPR pairs that can be extracted from the wavefunction. In QFT, for some region $A$ and its complement, this number is infinite; however, if we think about separated subregions $A,B$ in the QFT, it may be finite. Quantifying such entanglement is complicated by the fact that the $AB$ system is now a mixed state, and so we must confront mixed-state entanglement measures. This operational approach has been successfully applied to QFT in \cite{Hollands:2017dov}. These results are again UV insensitive. We discuss mixed-state measures of entanglement further in \S\ref{sec:mixedstate}.

\subsection{Entanglement entropy: methods}
\label{sec:methods}

A major area of progress in the past decade  has been in the refinement of traditional techniques and the development of new methods to compute entanglement entropy in QFT. 

Path integral techniques have proven powerful for interacting CFTs. One computes first the R\'enyi entropies,
\begin{equation}
S_n(A) = \frac{1}{1-n} \ln {\rm Tr} \rho_A^n
\end{equation}
for integer $n>1$. Analytically continuing in $n$ and taking the limit $n \to 1$ then give access to the von Neumann entropy. For special states and regions, e.g.\  a $(d-1)$-dimensional ball region for the vacuum of a $d$-dimensional CFT, symmetries of the relativistic (or conformally invariant) QFT can be used to compute the path integral for ${\rm Tr} \rho_A^n$ for all $n$. 
In \cite{Casini:2011kv} this was  utilized to compute entanglement in general CFTs, which was  successfully compared to the AdS/CFT prediction.  The origin of these symmetry methods lies in the Unruh effect \cite{Unruh:1976db}, and the abstract Bisagnono-Wichman theorem \cite{Bisognano:1976za}, which were originally discussed for $A$ being half of a Cauchy slice in Minkowski spacetime. Generalizations of this theorem to ball regions have been known for some time \cite{Hislop:1981uh}, but applications to entanglement entropy are relatively recent. Various supersymmetric techniques have been applied to this problem, see for example \cite{Nishioka:2013haa}. The path integral approach also allows for simple generalizations, such as perturbation theory for deformations of the state \cite{Rosenhaus:2014woa,Faulkner:2017tkh,Sarosi:2017rsq} and the shape \cite{Faulkner:2015csl,Faulkner:2014jva,Balakrishnan:2020lbp}, as well as a much larger class of examples in 2d \cite{Cardy:2016fqc}. 

For more general shapes and states, one needs to compute the path integral for the QFT residing on a singular manifold $\mathcal{M}_n$. This technique, referred to as the replica trick,  can be used to compute integer $n$ Renyi entropies. Powerful CFT techniques can be used to evaluate the functional integral. For instance, in 2d CFTs, one can compute entanglement entropy for various states and regions, and compare the results with holographic ones \cite{Faulkner:2013yia}. Alternately, the singular manifold can be dispensed with in favor of correlation function of twist operators in an orbifold of $n$ copies of the CFT. In 2d these are pointlike operators, so CFT methods, including bootstrap ideas, can be applied \cite{Headrick:2010zt,Hartman:2013mia,Headrick:2015gba,deBoer:2014sna,Chen:2015usa}. These methods apply to static and dynamical settings, and are particularly powerful for theories with holographic duals \cite{Hartman:2013qma,Asplund:2015eha}.  In higher dimensions, formulating the replica trick as a defect correlation function, where the boundary of the entanglement region $A$ contains a codimension-2 defect twist operator, has proven useful \cite{Bianchi:2015liz,Balakrishnan:2017bjg,Bianchi:2016xvf,Balakrishnan:2016ttg,Balakrishnan:2019gxl}. Deformations of the state and shape can then be computed using defect correlation functions and OPEs.

A particularly tricky issue with the replica trick is the continuation of $n$ away from the integers (so we can take $n \rightarrow 1$). There are several ways to approach this problem. One often applies Carlson's theorem to prove the existence of a unique analytic extension away from the integer values of $n$. The methods of moments has also been applied \cite{Dong:2021clv}. Analytic continuation from $n = 1/k$ for integer $k$ arises in stringy computations \cite{Dabholkar:1994ai,Witten:2018xfj}.  In gravitational theories the analytic continuation is often ``obvious'' \cite{Lewkowycz:2013nqa}, however the target level of rigor for a given paper drops often once the replica trick is invoked. Indeed, there are some known exceptions where the ``obvious'' answer is wrong \cite{Akers:2020pmf}. Some alternative methods have been explored in \cite{Agon:2013iva,DHoker:2020bcv}.

Finally, free theories have been analyzed in great detail, since the density matrix remains Gaussian under the partial trace. These methods go under the name of the correlation-matrix technique~\cite{Peschel_2003,Casini:2009sr}, since the correlation function of the fundamental fields, 
say $\langle\phi(x)\phi(y)\rangle$, can be viewed as a matrix whose indices are the positions $x,y$ and whose spectrum determines the entanglement entropy. These methods have been extended to compute the reflected entropy (see \S\ref{sec:mixedstate}) \cite{Bueno:2020vnx}, and work in both static and dynamic settings \cite{Cotler:2016acd}.

In parallel with these developments, new methods have been found by the algebraic QFT community to compute entanglement-like quantities. Most progress has come in the setting of free QFTs \cite{Longo:2017mbg,Ciolli:2019mjo}, where these arguments can be understood as rigorous versions of the correlation-matrix technique. More general results beyond free theories are however possible, including results for general superselection sectors \cite{Longo:2018obd} and an interesting class of states constructed using Connes cocycle flow \cite{Ceyhan:2018zfg}. 

Complementing these analytic methods is a range of numerical ones, mainly explored by the condensed-matter theory community. We will touch on some of these methods in  \S\ref{sec:simulation}.

As interesting physical problems arise, \emph{new techniques for computing quantum information quantities will likely need to be developed.} For example, one-shot entropies related to information processing tasks where only a single copy of a state is available (instead of many identical copies) have found applications in QFT and QG, e.g.~\cite{Czech_2015_length,Akers_2021_qes}.
\emph{Can we develop general tools to compute one-shot entropies and conditional entropies in QFT?}
It would also be interesting to develop, \emph{new methods for extracting the entanglement entropies from integer R\'enyis. Is it possible to use one-shot entropies for this purpose?}

\subsection{Entanglement in holography, gravity, and string theory}
\label{sec:hent}
 
A major spur to understanding entanglement entropies in QFTs, and source of data about them, has been the Ryu-Takayanagi (RT) formula \cite{Ryu:2006bv,Ryu:2006ef} and its various generalizations. The RT formula applies to holographic theories, which involve a large number of strongly coupled fields. It may seem paradoxical that we can relatively easily compute a difficult quantity like the entanglement entropy in such complicated theories, but we have by now become accustomed to the idea that holography simplifies the calculation of many quantities of physical interest.

Specifically, the von Neumann entropy of a region in a large $N$ gauge theory, or large central charge CFT, admits an asymptotic  expansion, $S = \sum_{g=0}^\infty\,  S_g(\lambda)\, N^{2-2g}$, with $\lambda$ being the effective 't Hooft coupling parameter of the theory. At large $N$ and large $\lambda$, the RT formula~\cite{Ryu:2006bv} posits that  the strong coupling result for the von Neumann entropy is geometrized in terms of the area of a minimal surface homologous to $A$ (and therefore anchored on the entangling surface $\partial A$):
\begin{equation}\label{eq:RT}
\lim_{\lambda \to \infty} \, S_0(\lambda) = \text{min}_X \, \frac{\text{Area}(\mathcal{E}_A )}{4\,G_N}  \,,\qquad 
X =\left\{\mathcal{E}_A \; : 
\exists \;\mathcal{R}_A \subseteq \mathcal{M} \text{ s.t. } \partial\mathcal{R}_A =
 \mathcal{E}_A  \cup A  
 \right\}.
\end{equation}  
The RT formula applies to static, or more generally time reversal-invariant, bulk spacetimes; in the above formula, $\mathcal{M}$ refers to the bulk Cauchy slice invariant under the time reflection. The RT formula can also be expressed in terms of so-called \emph{bit threads}, Planck-scale bulk curves connecting $A$ and its complement and representing Bell pairs in a distillation of the state \cite{Freedman:2016zud}. In the covariant Hubeny-Rangamani-Takayanagi (HRT) formula \cite{Hubeny:2007xt}, the minimization is over \emph{extremal} spacelike codimension-2 surfaces homologous to $A$. These prescriptions were justified using gravitational replicas in \cite{Lewkowycz:2013nqa} and \cite{Dong:2016hjy}, respectively. It can also be usefully reformulated as a \emph{maximin} prescription, in which the area is minimized within a Cauchy slice and then maximized over the choice of Cauchy slice \cite{Wall:2012uf}.

It was shown early on that the RT and HRT formulas obey the crucial SSA inequality \eqref{SSA} \cite{Headrick:2007km,Wall:2012uf,Headrick:2013zda}. This is necessary for the consistency of the formula; however, the proof is remarkably simple --- far simpler than the quantum proof of SSA --- and it remains unknown what feature of holographic states allows for such a simple proof. A slight generalization of the proof implies that the tripartite information \eqref{I3def} is non-positive in holographic states \cite{Hayden:2011ag,Headrick:2013zda}; unlike SSA, this inequality (called monogamy of mutual information, MMI, in the holographic literature) is not a general property of quantum states. An understanding of this property in terms of bit threads and bulk locality was proposed in \cite{Hubeny:2018bri}. A different explanation in terms of an ansatz for the entanglement structure of holographic states was put forward in \cite{Cui:2018dyq}, but this explanation was contested in \cite{Akers:2019gcv}, and the situation remains unclear. One implication of MMI is that the HRT formula obeys all members of the infinite family of conditional entropy inequalities discovered by Cadney-Linden-Winter \cite{LindenWinter2005,Cadney2011}, establishing that it obeys \emph{all} known general properties of the von Neumann entropy. It was shown in \cite{Bao:2015bfa} that the RT formula in fact obeys an infinite set of further inequalities that, like MMI, are not obeyed by general quantum states. The so-called holographic entropy cone defined by these inequalities has been extensively studied \cite{Cui:2018dyq,He:2019ttu,He:2020xuo,Hubeny:2018ijt,Hubeny:2018bri,Hubeny:2018trv}. Despite some evidence in the affirmative \cite{Erdmenger:2017gdk,Caginalp:2019mgu,Czech:2019lps}, the question of \emph{whether these inequalities are also obeyed by the HRT formula, as well as their significance in terms of the structure of holographic states}, remains open.

The proof that the HRT formula obeys SSA makes use of the Einstein equation and null energy condition, an illustration of the remarkable relationship between spacetime geometry and quantum information theory brought to light by holographic dualities. This relationship also works in the other direction: it is possible to derive the Einstein equation (perturbatively about AdS) from the HRT formula and properties of entanglement entropies (specifically, the first law of entanglement) \cite{Lashkari:2013koa,Faulkner:2013ica,Sarosi:2017rsq,Faulkner:2017tkh}.

The classical gravity answer \eqref{eq:RT} has two sources of corrections: quantum gravitational effects, which are both perturbative and non-perturbative in $1/N$; and string corrections, which are both perturbative and non-perturbative in $1/\lambda$. In $1/\lambda$ perturbation theory, classical string corrections are understood. For this purpose, it suffices to write down the higher derivative modifications to the Einstein-Hilbert dynamics that arise from the worldsheet string corrections \cite{Gross:1986iv}.  Within this effective field theory, modifications to the holographic entanglement entropy formulae can be analyzed \cite{Hung:2011xb,deBoer:2011wk,Dong:2013qoa,Camps:2013zua,Dong:2017xht}. The resulting expressions can be viewed as generalizations of the Wald entropy for higher derivative theories of gravity \cite{Wald:1993nt}.  For stationary black holes with bifurcate Killing horizons,  the entropy is the integral of a local functional, the Noether charge,  over the codimension-2 bifurcation surface (whose extrinsic curvature tensor vanishes identically).  In contrast,  the generalization of the extremal surface has non-vanishing extrinsic curvature and the resulting expression includes such contributions. This higher curvature corrected entropy formula also has the correct structure to give quasilocal entropy of dynamical horizons respecting the second law \cite{Wall:2015raa,Bhattacharyya:2016xfs,Bhattacharyya:2021jhr}.  

Finite $\lambda$ corrections are, however, as yet poorly understood. Essentially, \emph{one seeks an expression in closed string field theory that gives the genus zero contribution to the von Neumann entropy}. The prototype example is the thermofield double state, where we can ask how to derive the Bekenstein-Hawking entropy for black holes directly from the string worldsheet. Early attempts to tackle this question are \cite{Susskind:1994sm,Dabholkar:1994ai}, while recent attempts motivated by entanglement include 
\cite{He:2014gva,Balasubramanian:2018axm,Naseer:2020lwr}.  Thus far,  no clean derivation exists. The technical reason for this state of affairs owes to two issues. On the one hand, the tree level (genus zero) partition function of the string naively vanishes by worldsheet conformal invariance; on the other hand, the standard replica construction appears intractable from the worldsheet perspective \cite{Witten:2018xfj} (even for open strings). The first issue could  potentially be overcome  using suitable analytic continuation tricks, such as those employed in the context of Liouville theory to compute sphere and disk partition function or low-point correlators \cite{Dorn:1994xn,Erbin:2019uiz,Mahajan:2021nsd,Eberhardt:2021ynh}.   
Attempts have been made to understand this question in the context of quasi-topological field theories, e.g.\ 2d large $N$ Yang-Mills \cite{Donnelly:2016jet,Donnelly:2019zde}, and within the topological open/closed string duality \cite{Hubeny:2019bje,Donnelly:2020teo,Jiang:2020cqo}.  An analysis of partition functions of weakly coupled CFT duals from the dual string perspective (in the AdS$_3$/CFT$_2$ context), appears to suggest that worldsheet does not preferentially pick out a particular target space geometry in the semiclassical (large $N$) limit \cite{Eberhardt:2021jvj}. This is an important open question.

Quantum gravitational corrections at the leading semiclassical order were understood initially in \cite{Faulkner:2013ana} as contributions from the entanglement of bulk degrees of freedom in the homology region $\mathcal{R}_A$. This statement led to the quantum extremal surface proposal \cite{Engelhardt:2014gca}  (justified in \cite{Dong:2017xht}), which has played an important role in recent developments in the black hole information problem. This  prescription  can be shown to arise from the presence of replica wormhole saddles in the Euclidean quantum gravity path integral \cite{Almheiri:2019qdq,Penington:2019kki}. These derivations  have worked in a regime where the quantum corrections are amplified and made comparable to the tree level result, which enables one to discern the change in the nature of saddle point configurations. A complete understanding, however, requires \emph{defining the von Neumann entropy for subregions in theories with gravitational dynamics}. Attempts in this direction 
at the perturbative level, in particular dealing with the edge modes of the gravitational field, include \cite{Camps:2018wjf,Donnelly:2020xgu,Benedetti:2019uej,David:2020mls,David:2022jfd,Benedetti:2021lxj}. Much more about recent developments in the black hole information problem may be found in the white paper devoted to that topic \cite{Bousso:2022ntt}.


    \subsection{Mixed-state entanglement and correlation}
    \label{sec:mixedstate}

As noted above, the von Neumann entropy of a subsystem $A$ is a measure of the amount of entanglement between $A$ and the rest of the system $A^c$, assuming that the full system is in a pure state. Often, however, one is interested in correlations and entanglement between two subsystems $A$ and $B$ --- for example two spatial regions in a field theory --- that together do not make up the whole system, or between $A$ and $A^c$ when the full system is in a mixed state, such as a thermal state. These two scenarios are closely related, since, even if the full system is pure, if $AB$ does not make up the full system, then $\rho_{AB}$ will typically be mixed; conversely, if $\rho_{AA^c}$ is mixed, it can be purified by adding a third subsystem $O$, so that $\rho_{AA^cO}$ is pure, and in that point of view $AA^c$ is no longer the full system.

We have already mentioned one measure of correlation, namely the mutual information $I(A:B)$. This has many good properties: it is non-negative, is zero if and only if the state factorizes, and (by SSA) is monotonic under inclusion of more subsystems. It also has the virtue of being calculable, to the extent that one can calculate von Neumann entropies. In a field theory, if $A$ and $B$ are separate regions (not sharing a common boundary) then $I(A:B)$ is UV-finite and regulator-independent, and bounds correlation functions of bounded operators \cite{PhysRevLett.100.070502}:
\be
\left(\frac{\langle \mathcal{O}_A\mathcal{O}_B\rangle_{\text{conn}}}{\|\mathcal{O}_A\|\,\|\mathcal{O}_B\|}\right)^2\le2I(A:B)\,.
\ee

The mutual information quantifies the total amount of correlation, including both entanglement and classical correlation. Often one is interested in knowing just the amount of entanglement between $A$ and $B$. Unless $\rho_{AB}$ is pure, however, this is not a well-defined question: there is no single quantity that captures the ``amount of entanglement''. Rather, within a mixed state on $AB$ there are several different kinds of entanglement and correspondingly many distinct measures, which are useful for different purposes; see \cite{Plenio:2007zz} for a review. These measures are often defined in terms of an optimization problem; for example, the entanglement of formation is the number of Bell pairs required to prepare $\rho_{AB}$ under LOCC (local operations and classical communication), while the entanglement of distillation is the number of Bell pairs that can be produced from $\rho_{AB}$ using LOCC. There are also correlation measures that weight the entanglement and classical correlation differently and can therefore be used to quantify the amount of entanglement; again, these will not agree with each other. While this is a rich subject in quantum information theory, it is an unfortunate fact that very few of these measures are calculable in practice in quantum field theories.

A simple example of an entanglement measure that is calculable (again, assuming one can calculate von Neumann entropies) is the conditional information, $H(A|B)=S(AB)-S(B)$. This is non-negative for separable states, so if it is negative then entanglement is present. (However, the converse does not hold.) Over the past few years, there has been significant progress in finding other entanglement and correlation measures that are calculable in field theories. The first of these was the logarithmic negativity, defined as $\ln\text{tr}|\tilde\rho_{AB}|$, where $\tilde\rho_{AB}$ is the partial transpose of $\rho_{AB}$ (i.e.\ $\rho_{AB}$ transposed just on the $A$ indices) \cite{Simon:1999lfr,Peres:1996dw,Horodecki:1996nc,Zyczkowski:1998yd,Eisert:1998pz,Vidal:2002zz,Plenio:2005cwa}. If this quantity is positive (or more generally if $\tilde\rho_{AB}$ has negative eigenvalues) then entanglement is present. The logarithmic negativity can be calculated via a replica trick, and has been calculated in a variety of spin systems and field theories \cite{Calabrese:2012ew,Calabrese:2012nk,Calabrese:2014yza,Kulaxizi:2014nma,Rangamani:2015qwa,Ruggiero:2016yjt,Mbeng:2016fnc,Malvimat:2017yaj,Shapourian:2019xfi,Shapourian:2020mkc}. In holography, negativity can be computed for ball-shaped regions of a CFT \cite{Rangamani:2014ywa}, but there is no general prescription. Various groups have attempted to study this using tensor network toy models \cite{Kudler-Flam:2021efr,Dong:2021clv,Vardhan:2021npf,Vardhan:2021mdy}, but these have not led to a clear geometric picture; despite some work on the problem (e.g.\ \cite{Chaturvedi:2016rft,Chaturvedi:2016rcn,Kusuki:2019zsp}), it remains unclear whether any such picture exists.

Two other correlation measures have received much attention, due in part to their conjectured holographic dual: the entanglement of purification (EOP) and the reflected entropy. The EOP is defined as the minimal value of $S(A\tilde A)$ among purifications $\rho_{AB\tilde A\tilde B}$ of $\rho_{AB}$. Again, since this quantity is defined via an optimization, it can be hard to compute except in simple systems, but it can be estimated or bounded in various ways \cite{Takayanagi:2017knl,Nguyen:2017yqw,Bhattacharyya:2018sbw,Caputa:2018xuf,Bhattacharyya:2019tsi,Camargo:2020yfv}. Like the mutual information, it is non-negative, vanishes precisely on factorized states, and is monotonic under inclusion; in that sense it is a good correlation measure. It takes account of both entanglement and classical correlation, but weights them differently than the mutual information, so by comparing the two quantities one can learn about the relative amounts of these two types of correlation in a given state.

The reflected entropy is similarly defined as $S(A\tilde A)$ under a purification $\rho_{AB\tilde A\tilde B}$ of $\rho_{AB}$, but, in contrast to the EOP, the purification is not optimized over but rather fixed to be the canonical one (defined similarly to the thermofield double state vis-\`a-vis the thermal state) \cite{Dutta:2019gen}. Like the EOP, the reflected entropy can be used to study the relative amounts of entanglement and classical correlation in a given bipartite system. The fact that its definition does not involve an optimization makes this quantity much easier to calculate than the EOP; methods include the replica trick and correlation-matrix technique. On the other hand, it is harder to investigate its general properties; for example, it is not known whether the reflected entropy is monotonic under inclusion, and in that sense constitutes a good correlation measure. 

Using a replica construction, \cite{Dutta:2019gen} argued that the reflected entropy is dual to the entanglement wedge cross-section, defined as the minimal surface separating $A$ and $B$ within their joint entanglement wedge \cite{Takayanagi:2017knl,Nguyen:2017yqw}. Interestingly, the EOP was also argued to share the same holographic dual \cite{Takayanagi:2017knl,Nguyen:2017yqw}. While currently there is better evidence for the reflected entropy to be dual to the entanglement wedge cross-section, it may be that both proposals are correct, and in holographic states the two quantities are simply equal --- not an implausible idea, given that such states are already known to have a very special entanglement structure. An important open problem is thus to determine \emph{whether the entanglement wedge cross-section is equal to the EOP, the reflected entropy, both, or some other easily characterized quantity in the field theory}.

Even outside of holography, the reflected entropy is potentially a very useful correlation measure in QFT and even in general quantum systems, thanks to its calculability. This is an example of investigations in QFT and quantum gravity paying back dividends to quantum information science. However, it is important to better understand the fundamental properties of reflected entropy. Key questions include: \emph{Is the reflected entropy is monotonic under inclusion, thereby qualifying it as a valid correlation measure? What is the exact connection between modular nuclearity \cite{buchholz1990nuclear}, the split property, and reflected entropy? And what are sufficient conditions for reflected entropy to be finite?} (See \cite{Panebianco:2021vaz}.)

    \subsection{Complexity}
    \label{sec:complexity1}

Thus far we have been discussing various aspects of the entanglement and correlation between subsystems, such as spatial regions in a QFT, as measured by entropies and related quantities. We can get further information by asking how hard it is to generate a pattern of entanglement and what we can do with it. For example, for an extended quantum system, such as a spin system or a lattice-regularized QFT, one could ask how many operations of a given type are required to reach the ground state starting from a completely unentangled state. If the operations in question are quasilocal --- for example, they act on at most two lattice sites at a time --- then producing a highly entangled state may require a large number of operations. Since we can think of the operations as gates in a quantum circuit, the number of operations required is called the \emph{circuit complexity} of the state. It is related to the tensor network description of states, to be discussed in section \ref{sec:simulation} below. Since it is defined in terms of an optimization, the circuit complexity is an intrinsically difficult quantity to calculate. Moreover, in addition to the usual dependence on the regulator for a QFT, it depends on the choice of initial state and gate set, so there is no single well-defined quantity called \emph{the} circuit complexity. Despite these difficulties, there are reasons to believe that this notion captures interesting features of field-theory states that are inaccessible to entropy-based measures.

Like so many applications of quantum information to QFT, the circuit complexity story was initially motivated from holography. As we will review in \S \ref{sec:complexity2}, the circuit complexity of holographic states has been conjectured to be related to either the volume of maximal slices or the action of Wheeler-de Witt patches in the bulk anchored to a given time-slice \cite{Susskind:2014rva,Stanford:2014jda,Brown:2015bva}. Since then, however, the notion has taken on a life of its own, and despite the difficulty in calculating it, has been studied extensively in free and weakly-coupled field theories, spin chains, and so on 
\cite{Camargo:2020yfv,Jefferson:2017sdb,Hackl:2018ptj,Guo:2018kzl,Chapman:2018hou,Caceres:2019pgf,Agon:2018zso,Balasubramanian:2021mxo,Bhattacharyya:2018bbv,Balasubramanian:2019wgd,Khan:2018rzm}.

Various potentially useful notions of complexity are closely related to the circuit complexity of a pure state. For example, one can define the circuit complexity of an operator as the minimum number of gates required to build the operator \cite{Roberts:2018mnp,Qi:2018bje,Parker:2018yvk,Mousatov:2019xmc,Gomez:2019xrl,Magan:2020iac,Flory:2020eot,Susskind:2020gnl,Jian:2020qpp,Kar:2021nbm}. One can also define the complexity of a state in terms of a continuous distance function on the Hilbert space, which can for example be the geodesic distance with respect to some metric \cite{Belin:2018bpg,Chapman:2017rqy}. One can also define the complexity of a state in terms of the number of sources required to prepare it using a path integral \cite{Caputa:2017urj,Caputa:2017yrh,Takayanagi:2018pml,Skenderis:2008dg,Botta-Cantcheff:2015sav,Marolf:2017kvq}. Finally, the notion of state complexity can be expanded in various ways to include subsystem states or more generally mixed states \cite{Agon:2018zso,Caceres:2019pgf}. All of these notions are under active investigation, and it remains an open problem \emph{to define those that are most useful for classifying field theories and understanding their time evolution}, as well as \emph{to develop tractable methods for computing them}. 
We can expect that the resulting new ideas for defining and computing various kinds of complexity will be of interest beyond the QFT setting, providing another example of a dividend paid back to quantum information science from fundamental physics.

\section{Symmetry, topology, and renormalization group flows}
\label{sec:symtop}

We now turn to how quantum information ideas interplay with some key non-perturbative structures in QFT. We first discuss symmetries in QFT and gravity. Then we consider renormalization group (RG) flows, including how entropy inequalities constrain such flows, and how IR fixed points can be classified by their patterns of entanglement. Finally, we highlight the special case of IR fixed points corresponding to topological QFTs, where entanglement and topology become intertwined.

 \subsection{Symmetries, charges, and superselection sectors}
\label{sec:supersel} 

Symmetries in quantum field theory can also be studied from a quantum information perspective. The statistical properties of the charge fluctuations in localized regions can be extracted in various ways.

Symmetry resolved entanglement \cite{Goldstein:2017bua}
studies the different charged sectors of the reduced density matrix that arises for a theory with a local symmetry and a state that is invariant under that symmetry. In the path integral formulation, such sectors can be studied by inserting flux at the twist operator, or equivalently by turning on a chemical potential for the charge.  Correspondingly, the holographic description (at least for highly symmetric regions) is determined by a charged hyperbolic black hole. This was studied in \cite{Belin:2013uta,Dowker:2015dpd}. There is an interesting application of symmetry resolved entanglement to black hole physics in \cite{Milekhin:2021lmq}. In particular, it provides an interesting probe of the non-existence of global symmetries in quantum gravity.

In the algebraic approach, superselection sectors govern the global charges of a quantum field theory \cite{haag2012local}. These charges leave their imprint on algebras associated to disjoint regions, via non-local operators that violate additivity or Haag duality of these algebras \cite{Casini:2019kex}. Such properties of local algebras are not necessarily required by causality and consistency under restriction to smaller regions. Order parameters based on the relative entropy have been developed to probe these superselection sectors \cite{Casini:2020rgj,Magan:2020ake};
see also \cite{Xu:2018uxc,Hollands:2020owv}.  
Generalized global symmetries can also be studied from this perspective \cite{Casini:2021zgr}. 

From the algebraic perspective, the short-distance fluctuations of charges for disconnected regions that approach each other have a universal form for the various charged sectors \cite{Casini:2019kex}. This universal form is in turn related to the form of the high temperature density of states of the theory defined on a compact space. This form was conjecture in \cite{Harlow:2021trr} and proven in \cite{Magan:2021myk}. It is also related to the equipartition property \cite{Xavier:2018kqb} for symmetry resolved entanglement.

\subsection{RG flows and c-theorems}
\label{sec:RG}

Our understanding of effective field theories, pioneered by Wilson \cite{Wilson:1973jj,Wilson:1974mb} and made precise by Polchinski \cite{Polchinski:1983gv},   relies on the fact that dynamics below some cut-off scale is only sensitive to a handful of relevant parameters. From a microscopic viewpoint, attaining this low energy effective description, relies on tracing out high energy degrees of freedom, an intrinsically lossy process. Intuitively, therefore, one anticipates the existence of a measure of the number of degrees of freedom that is monotone under the renormalization group. 

In two dimensions, exploiting details of energy-momentum tensor correlations, Zamolodchikov  \cite{Zamolodchikov:1986gt} proved a $c$-theorem, exhibiting a function of the couplings $C(g_i)$ that is monotonically decreasing as a function of scale and attains a stationary value equal to the central charge at fixed points.  The search for analogous statements in higher dimensions led to two interesting statements: an $F$-theorem for three-dimensional theories \cite{Jafferis:2011zi} (see \cite{Myers:2010xs,Myers:2010tj}) and $a$-theorem in four dimensions \cite{Komargodski:2011vj}. The latter was proven with the understanding of an effective description of softly broken conformal symmetry, while the former was motivated by the understanding of supersymmetric RG flows and localization. 

While the conventional field theory methods rely on dimension-specific techniques, an interesting overarching principle can be discerned by examining the entropic proofs of the statements. Specifically, using the fact that the vacuum of a CFT is a Markov state, viz.,  the fact that SSA is saturated for a pair of subregions with boundaries lying on the light-cone, one finds a universal proof for the irreversibility of the RG flow \cite{Casini:2017vbe}. Historically, the two-dimensional $c$-theorem was proved by Casini-Huerta using SSA in \cite{Casini:2004bw} (reviewed in \cite{Casini:2006es}). A clever adaptation of this technique, with inspiration from holographic explorations \cite{Liu:2012eea},  led these authors to the first proof of the $F$-theorem in three dimensions \cite{Casini:2012ei}. The $a$-theorem was understood more recently, building on the understanding of operator algebras and the structure of the modular Hamiltonian on null planes \cite{Casini:2017roe,Casini:2017vbe}. 

An interesting avenue for further developments is to ascertain \emph{whether analogous monotone functions exist for higher-dimensional field theories (at least in five and six dimensions where superconformal critical points exist) and how one can understand them in terms of information-theoretic data}. 

The discussion of RG is typically anchored to the vacuum dynamics. Inspired by holography, it is interesting to enquire about state-dependent features of renormalization. For instance, one would anticipate that the process of integrating out high-energy degrees of freedom leads to a quantum channel for the low-energy density operator. While there is no general statement to this effect to date, there have been recent efforts to understand real-space RG in terms of quantum error correction \cite{Furuya:2020tzv,Furuya:2021lgx} which may provide a pathway to this question. A related issue is the nature of effective field theories for open quantum systems; see \S\ref{sec:openqs}.

    \subsection{Classification of phases}

Entanglement and complexity also play an important role in the classification of IR fixed points of RG flows. In the context of many-body physics, both in condensed matter examples and engineered systems built from atoms, molecules, and photons,  
such fixed points describe distinct phases of matter. Classification of these phases is a major challenge in many-body physics, intimately tied to the classification of QFTs.

Systems with one spatial dimension ($d=2$) provide a wealth of instructive examples. In the absence of any symmetry and assuming a mass gap, it is believed that there is a unique IR fixed point --- the empty theory. There is no analog of the topological field theories (discussed below) in $d\geq 3$ dimensions, which can have a degenerate space of vacuum states. However, when the system has a symmetry, distinct gapped IR fixed points are sometimes possible~\cite{Fidkowski_2011,PhysRevB.83.035107,Turner_2011}. These are labelled by projective representations of the group and correspond to the physical possibility of protected edge states when the theory is defined on an interval. Remarkably, this structure was first discovered using entanglement and tensor network ideas: such 2d gapped IR fixed points should have a matrix product state representation, and these can be classified via projective representations of the symmetry acting on the virtual indices.

This classification has now been greatly extended, with many new and deep connections uncovered between anomalies, symmetries, and patterns of entanglement (e.g.~\cite{Chen_2013,Vishwanath_2013,Chong_2013,Burnell_2014,Wen_2013,freed2014relative,Kapustin_2014,Barkeshli_2020}). The result is a rich zoo of IR fixed points (in the field theory language) and corresponding set of phases of matter (in the many-body physics language). One interesting manifestation of this classification is in terms of quantum state preparation. Consider a lattice-regulated field theory, and suppose we want to prepare the ground state of this theory. When the IR fixed point describing the continuum limit is the trivial theory, all excitations in the system are heavy and there is no interesting IR structure in the quantum state. In this case, a finite-depth quantum circuit suffices to give a good approximation to the ground state. On the other hand, when the IR fixed point corresponds to a non-trivial symmetry-protected phase, then the corresponding state preparation circuit cannot have finite depth if it is also required to be symmetric~\cite{Huang_2015}. Such theories are known as invertible QFTs (see e.g.~\cite{freed2014anomalies}).

An even richer setting occurs when we consider non-invertible QFTs, which correspond to topological field theories with degenerate ground states on a spatial manifold of non-trivial topology. Chern-Simons theory in three dimensions, which is an exactly solvable  topological field theory  \cite{Witten:1988hf}, is a useful playground for analyzing this physics, including topological entanglement entropy \cite{Kitaev:2005dm,Levin:2006zz}.  One can use the replica method to obtain spatially ordered entanglement in this theory \cite{Dong:2008ft}. The result is an  entanglement spectrum determined by the quantum dimension \cite{Kitaev:2005dm} and the number of connected components of the entangling surface. For example, for Chern-Simons on $S^3$, the entanglement entropy for a spatial bipartitioning gives all R\'enyi and von Neumann entropies simply equal to $\log Z_\text{CS}(S^3)$. One can extract more interesting information by considering states involving Wilson lines along various knots and links as discussed in \cite{Dong:2008ft, Balasubramanian:2016sro,Salton:2016qpp,Balasubramanian:2018por}, while \cite{Wen:2016bla} considered negativity in these theories. One can understand the edge modes in these systems quite cleanly \cite{Elitzur:1989nr,Das:2015oha,Wen:2016snr}. Relations between  this topological entanglement and BTZ black hole entropy have also been proposed \cite{McGough:2013gka}, and as noted in \S\ref{sec:hent} one can also relate Chern-Simons entanglement to topological string theory. Finally, such theories must also have high state complexity, since one can show that the topological S-matrix is invariant under finite-depth unitary transformations~\cite{Haah_2016}.

Chern-Simons theories with dynamical matter \cite{Gaiotto:2007qi,Aharony:2011jz,Giombi:2011kc} have interesting applications. They naturally appear as theories on M2-branes \cite{Aharony:2008ug,Aharony:2008gk} and  exhibit Bose-Fermi duality \cite{Jain:2013gza,Hsin:2016blu}. Furthermore, they have an interesting phase structure at large $N$ \cite{Jain:2013py,Aharony:2012ns}, providing examples that interpolate between vector- and matrix-like models \cite{Chang:2012kt}. The former are dual to higher spin gravity at large $N$, while the latter lead to more familiar gravitational duals. The fractional statistics of these theories is manifested by a deformation of the thermal distribution functions \cite{Geracie:2015drf,Minwalla:2022sef}. Information-theoretic properties of these theories have not been explored thus far; such studies could provide useful insights both into field theory dynamics and into the holographic map. 

A more exotic example is provided by so-called fracton phases of matter~\cite{Haah_2011,Vijay_2015,Pretko_2017}. These are novel phases realized in solvable many-body models which do not correspond to any conventional IR fixed point of a quantum field theory. A prototypical example is a higher spin gauge theory (without Lorentz symmetry) where conservation of dipole moment reduces the mobility of charged excitations~\cite{Pretko_2017}. The exotic physics of such systems, which host novel kinds of ground state entanglement and excitations with restricted mobility, has prompted the search for extended kinds of field theories that can describe these unusual phases of matter, e.g.~\cite{Seiberg_2020,ShuHeng_2020}. \textit{Continued development of such fractonic field theories is certainly needed, as well as work connecting the broad idea of generalized global symmetries to entanglement and complexity.}

There are many other open questions. For example, the classification of gapless phases is poorly understood. RG monotones, including many motivated by entanglement considerations, help to constrain the space of RG fixed points. However, \emph{a direct classification of these fixed points, e.g.\ via something like topological entanglement entropy, is still lacking}. In the case of systems at finite density, the situation is even more interesting, since qualitatively new kinds of finite terms appear in the entanglement~\cite{PhysRevLett.96.100503}.

\section{Dynamics}
\label{sec:dynamics}

In this section, we describe recent progress arising from the study of quantum information dynamics. First, we discuss the physics of information spreading and scrambling in QFT and quantum gravity, and the new connections being established with the field of many-body quantum chaos. Next, we discuss how entropic constraints have been used to prove very general energy inequalities in quantum field theory. Then we discuss open system dynamics of field theories. Finally, we return to the subject of complexity and discuss the conjectured relationship between complexity growth and the growth of black hole interiors.

\subsection{The spreading of information}

Under time evolution, quantum information tends to spread across many degrees of freedom. It spreads both spatially, as signals propagate locally through the system, and internally, among the degrees of freedom at each site. A remarkable discovery that has precursors in the 1970s but has come into focus over the last decade is that the spreading of quantum information obeys universal laws. In a wide variety of complex systems, including condensed matter, strongly interacting QFTs, and black holes, information spreads according to general principles and subject to fundamental bounds imposed by locality and by information-theoretic inequalities. These bounds play a key role in questions such as how locality emerges in quantum gravity and how to characterize the dynamics of strongly correlated materials.
 
The spread of quantum information can be characterized in several ways, and various types of information spread at different rates. In relativistic systems, of course, no signal can travel faster than the speed of light $c$. Lieb-Robinson proved that non-relativistic systems with a local Hamiltonian also obey a fundamental speed limit, $v_{LR}$, whose value depends on the microscopic details of the Hamiltonian \cite{Lieb:1972wy}.

Recently, several new measures of information spreading have been developed. The entanglement velocity $v_E$ is defined by a time derivative of the entanglement entropy, and interpreted as a rate at which entanglement spreads across space \cite{Hartman:2013qma,Liu:2013iza}; the butterfly velocity $v_B$ was originally defined by the spread of quantum chaos \cite{Shenker:2013pqa}, but often it also behaves as a state-dependent Lieb-Robinson velocity that constrains other dynamics \cite{Roberts:2016wdl}; the information velocity $v_I$ \cite{Mezei:2016wfz,Mezei:2016zxg,Couch:2019zni} measures the growing size of the region needed to recover the entanglement of a physical system with a reference system. All of these recent measures of information speed have been developed by combining the toolkits of quantum information theory, condensed matter theory, and high-energy theory, with many of the insights originating in the physics of black holes.

Calabrese-Cardy were the first to systematically study the spread of entanglement in quantum field theory \cite{Calabrese:2005in}, using the model of a quantum quench from a gapped to an ungapped Hamiltonian. They found that in 1+1-dimensional rational CFTs, entanglement spreads ballistically at the speed of light; thus $v_E = c$ in these systems. In fact, the ballistic spread of entanglement is nearly universal (see for example \cite{Amico:2007ag} for a review), and it has been observed experimentally in cold atoms \cite{cheneau2012light}. In the AdS/CFT correspondence, quantum quenches are holographically dual to black holes far from equilibrium \cite{Hubeny:2007xt,Hartman:2013qma,Liu:2013iza}, so the dynamics of black holes can be used to study the spread of entanglement in interacting quantum field theories in higher dimensions, and in non-rational CFTs in two dimensions \cite{Asplund:2015eha}. In higher dimensions, entanglement spreads more slowly; it satisfies  the inequalities \cite{Avery:2014dba,Hartman:2015apr,Casini:2015zua,Mezei:2016wfz,Mezei:2016zxg,Couch:2019zni}
\begin{align}\label{vbounds}
v_E \leq c , \quad v_E \leq v_{B} \ , \quad v_E \geq v_I \  .
\end{align}
These bounds follow from the monotonicity of relative entropy, applied to nested regions in spacetime. They can also be understood from a membrane model of entanglement propagation that has been developed through a combination of condensed matter methods \cite{Jonay:2018yei} and holography \cite{Mezei:2018jco}. Inequalities such as these also appear to be an important ingredient in understanding the microscopic origin of locality in quantum gravity, as manifested in the principle of entanglement wedge nesting in AdS/CFT \cite{Wall:2012uf,Headrick:2014cta}. 

There are many important unanswered questions that we expect to drive progress on this topic in the next decade. \emph{What are the other fundamental bounds on the spreading of quantum information? Can locality in quantum gravity be placed more firmly into the context of quantum information theory, and, in particular, quantum Shannon theory? Does this provide insight into the profound problem of understanding causality in non-perturbative quantum gravity, which underlies the black hole information paradox?} For recent ideas in each of these directions, see for example \cite{Casini:2004bw,Casini:2017vbe,Baumann:2019ghk,May:2021nrl}. 

Constraints on the spread of chaos and quantum information are also intimately tied to the theory of transport in strongly interacting systems. This is especially interesting because such bounds may ultimately be responsible for the unexplained phenomenon that a wide class of high-temperature superconductors exhibit $T$-linear resistivity in the ``strange metal'' regime above the superconducting transition. There is a longstanding conjecture that this universality results from an upper bound on the scattering rate of electrons, known as the Planckian bound \cite{sachdev1999quantum,damle1997nonzero,zaanen2004temperature},
\begin{align}
\tau \gtrsim \frac{\hbar}{k_B T} \ .
\end{align}
This famously leads to the conjecture of Kovtun-Son-Starinets that the ratio of viscosity to entropy density is bounded by $\eta/s \gtrsim \hbar/4\pi k_B$ 
and saturated by black holes \cite{Kovtun:2004de}. At present, there is no definitive link between information-theoretic bounds such as \eqref{vbounds} and the Planckian bound, but there are several hints. For example, the viscosity bound can be formulated more generally as a bound on the diffusion constant $D$ \cite{Hartnoll:2014lpa}, and in many examples, it takes the form 
$D \lesssim \hbar v_B^2 / k_B T$, where $v_B$ is the butterfly velocity \cite{Blake:2016wvh}. There are also a number of other recent bounds on transport derived from causality \cite{Hartman:2017hhp,Lucas:2017ibu}, the averaged null energy condition \cite{Delacretaz:2018cfk}, and other information-theoretic arguments. \emph{Can these ideas be developed into a more detailed theory of Planckian transport? What are the consequences of information-theoretic bounds for other transport phenomena in condensed matter?}

Such bounds are also interesting in disordered systems, since these systems often exhibit slow transport dynamics. The extreme limit of this class of behaviors is many-body localization (see~\cite{RevModPhys.91.021001} for a review), where transport is completely arrested, but interesting slow dynamics have been observed even if full localization is not achieved (and its fate in the thermodynamic limit is still contested, e.g.~\cite{sels2021thermalization}). There are many questions here, including \emph{whether sub-ballistic operator growth (vanishing $v_B$) implies sub-diffusive transport (vanishing $D$)} (e.g.~\cite{Nahum_2018,Yoo_2020}).

\subsection{Scrambling and chaos}

Hayden-Preskill \cite{Hayden:2007cs} formulated a circuit model of black hole dynamics in which the black hole Hamiltonian rapidly randomizes the quantum state. Sekino-Susskind \cite{Sekino:2008he} conjectured that in any quantum system, this scrambling process can occur no faster than $t_s \sim \frac{\beta}{2\pi}\log S$, with $S$ the entropy, and that this bound is saturated by black holes. It is now understood that scrambling is ubiquitous in chaotic quantum systems with many degrees of freedom, from the SYK model to black holes and large-$N$ CFTs.  

A far-reaching new perspective on scrambling, developed over the last decade, reformulates it in terms of out-of-time-order correlation functions (OTOCs) \cite{kitaevtalk,Shenker:2013pqa,Shenker:2013yza,Roberts:2014isa}. Schematically, one considers observables of the form
\begin{align}
G_{\rm OTO} &= \langle [A(0), B(t)]^2 \rangle \,.
\end{align}
OTOCs measure the effect of small perturbations at time $t=0$ on the Heisenberg operators at time $t$, and thereby probe the onset of quantum chaos. In large-$N$ systems, they behave (after appropriate normalization and regularization) as
\begin{align}
    G_{\rm OTO} \sim 1 - \frac{1}{N} e^{\lambda t} + \cdots
\end{align}
The second term is the harbinger of chaos, and $\lambda$ is the Lyapunov exponent. Maldacena-Shenker-Stanford proved that, under very general conditions \cite{Maldacena:2015waa},
\begin{align}
    \lambda \lesssim \frac{2\pi}{\beta} \ .
\end{align}
Chaos therefore becomes appreciable at time $t_s \gtrsim \frac{\beta}{2\pi} \log N$, in agreement with the fast scrambling conjecture of Sekino-Susskind. In many cases, scrambling as diagnosed by OTOCs occurs on the same timescale as scrambling defined by information-theoretic measures \cite{Mezei:2016wfz}, and in general, sums over OTOCs can be used to bound the mutual information \cite{Hosur:2015ylk}.  These results are also tied to emergent causality in the AdS/CFT correspondence \cite{Camanho:2014apa} and to bounds on CFTs from the conformal bootstrap \cite{Hartman:2015lfa,Hartman:2016lgu,Afkhami-Jeddi:2016ntf}.  

These developments have begun to form the basis for a deeper understanding of quantum chaos, and how it relates to the dynamics of quantum information, as well as more traditional observables in quantum field theory. We expect this to continue providing new insights. Specific questions to pursue in this area include: \emph{Can applications of random matrix theory to chaotic quantum systems, which have played a crucial role in AdS$_2$/CFT$_1$, be extended quantitatively to higher dimensions? What is the role of eigenstate thermalization in the dynamics of higher-dimensional black holes and wormholes?}

Another important question is whether qualitatively new features emerge at finite $N$ in spatially extended systems. This is in essence a different thermodynamic limit: large volume instead of large $N$. In many-body models, including random circuits~\cite{Nahum_2018_opgrowth} and spin chains~\cite{Xu_2019_mpo}, it was found that the OTOC expands ballistically in spacetime and the characteristic length scale of the wavefront grows with time. To explain, consider an OTOC of the local operators of the form $G_{OTO}(x,t) = \langle [A(0,0),B(x,t)]^2\rangle$ as a function of $x$. In one spatial dimension, the crossover of $G_{OTO}$ from zero to maximal occurs around $x=v_B t$ and takes place over a length scale $\ell$. Large $N$ models have $\ell \sim \text{const}$ whereas many-body models with finite $N$ typically have $\ell \sim \sqrt{t}$. It is sometimes useful to think in terms of a velocity dependent Lyapunov exponent $\lambda(v)$ where $v=x/t$~\cite{Khemani_2018_velocity}, in which case the butterfly velocity corresponds to $\lambda(v_B)=0$. Large $N$ models have $\partial_v \lambda(v_B) \neq 0$, but in the many-body models the ``diffusive broadening'' corresponds to $\partial_v\lambda(v_B) = 0$. This broadening appears to be universal in one spatial dimension ($d=2$) at finite $N$~\cite{Xu_2019_local}, but \textit{seeing this behavior in QFT and quantum gravity at finite $N$ is an outstanding challenge.}

 \subsection{Energy and entropy bounds}
 \label{sec:bounds}

Vacuum-subtracted energy density in quantum field theory can be negative due to quantum fluctuations. This negative energy can potentially give rise to acausal, or otherwise pathological, gravitational dynamics when coupling the QFT to gravity. For example, traversable wormholes might provide shortcuts between distant points. The Hawking black hole area theorem and Penrose singularity theorems rely on assumptions about non-existence of various forms of negative energy \cite{Hawking:1973uf}. It is thus important to find general constraints on such negative energy. Surprisingly, these constraints have been shown to arise from quantum information considerations applied directly to the quantum field theory without gravity \cite{Casini:2008cr,Wall:2011hj}. This story thus connects to the broader paradigm of gravity from quantum information. 
    
One example is the Bekenstein bound \cite{Bekenstein:1980jp}:
    \be
    S(A)_\rho-S(A)_{|0\rangle}\le\langle K\rangle_\rho-\langle K\rangle_{|0\rangle}\,,
    \ee
    where $A$ is a half-space in Minkowski space, $\rho$ is an arbitrary state, $|0\rangle$ is the vacuum, and $K$ is the vacuum modular Hamiltonian. This bound follows from positivity of the relative entropy \eqref{relentdef} \cite{Casini:2008cr}. Related inequalities were proven in \cite{Bousso:2014sda,Bousso:2014uxa}.
Another example is the averaged null energy condition (ANEC)
    \be
 \mathcal{E}_+ :=    \int_{-\infty}^\infty dx^+\,T_{++}\ge0\,,
    \ee
    derived from monotonicity of relative entropy \cite{Faulkner:2016mzt}. (The ANEC was proven independently without quantum information considerations in \cite{Hartman:2016lgu} and an argument was given in favor of the ANEC for holographic theories \cite{Kelly:2014mra}.) Yet another case is the quantum null energy condition (QNEC),
    \be
    \label{eq:qnec}
    \langle T_{kk}\rangle\ge\frac{\hbar}{2\pi a}S''[\Sigma]\,,
    \ee
where $a$ is an infinitesimal area element on an entangling surface $\Sigma$ and $S''[\Sigma]$ is the second derivative of the entanglement entropy with respect to variations of $\Sigma$ with area $a$ in the direction $k$. The QNEC was conjectured in \cite{Bousso:2015mna}, proven for free theories and AdS/CFT in \cite{Bousso:2015wca, Malik:2019dpg}, and finally shown to follow in general QFTs from the analyticity of modular flowed correlation functions \cite{Balakrishnan:2017bjg}. Another argument for the QNEC is that it reduces to the ANEC in a special family of states constructed using Connes co-cycle flow \cite{Ceyhan:2018zfg}. This proof of the QNEC can be formulated rigorously in  the algebraic approach to quantum information. Several extensions and applications of this co-cycle flow result  have been explored \cite{Bousso:2020yxi,Levine:2020upy}. The QNEC as stated in \eqref{eq:qnec} is actually saturated in interacting theories, as was argued for in AdS/CFT \cite{Leichenauer:2018obf} and in general \cite{Balakrishnan:2019gxl}. This distinction between free and interacting theories is similar to the distinction between the form of the energy correlation functions of a free theory versus an interacting theory \cite{Hofman:2008ar}. 
     
Energy correlation functions, originally studied by Hofman-Maldacena, are related to correlation functions of the ANEC operator $ \mathcal{E}_+$ in Minkowski spacetime. Positivity of these correlation functions puts important constraints on CFT and QFT data \cite{Hofman:2008ar}.  In this way, quantum information has been used to constrain QFT theory space. 
    
Traversable wormholes have been constructed \cite{Gao:2016bin,Maldacena:2017axo}, exploiting violations of the ANEC that are allowed for non-achronal null geodesics (null geodesics containing timelike-separated points). Examples were originally constructed in AdS/CFT using non-local double trace deformations \cite{Gao:2016bin,Maldacena:2018lmt}, and quickly generalized to more realistic theories of gravity \cite{Maldacena:2018gjk,Horowitz:2019hgb,Fu:2019vco}.  The specific form of null energy comes from a Casimir-like energy, familiar from CFTs on a cylinder. Such traversable wormholes do not provide shortcuts through spacetime, since the length of the wormhole is restricted to be larger than the distance between the two throats in the ambient spacetime \cite{Graham:2007va,Wall:2010jtc}. 
    
Important open questions in this area include the following: \emph{Is there a proof of the improved QNEC in 2d CFTs conjectured in \cite{Wall:2011kb}? Is there a connection between causality constraints leading to the ANEC \cite{Hartman:2016lgu} and the quantum information argument mentioned here? What is the physical meaning of the higher-spin QNEC proven in \cite{Balakrishnan:2019gxl}? Is there a rigorous algebraic proof of the saturation of the QNEC? Does the ANEC place bounds on the quark gluon plasma studied in heavy ion collisions? Can these energy conditions, as applied to the standard model, be experimentally tested? Is there a proof for general spacetimes of the self-consistent achronal ANEC \cite{Graham:2007va}, the generalized second law, and the quantum focusing conjecture \cite{Bousso:2015mna}?}

\subsection{Open quantum systems}
\label{sec:openqs}

A quantum system interacting with an external environment does not evolve unitarily. Rather, the combined system and environment degrees of freedom together evolve unitarily, which upon tracing out the environment indicates that the system by itself evolves as dictated by a quantum channel. This is the characteristic of an open quantum system. The broad question of interest is to determine the general rules to construct effective field theories for such open quantum systems.

 The basic paradigm for such open quantum dynamics was laid out by long ago by Feynman-Vernon \cite{Feynman:1963fq}. Since one has to describe the operation of a quantum channel, one can work directly with density matrices, or equivalently in terms of a  doubled set of degrees of freedom for the system corresponding to the state vectors and their conjugates. The novel feature is that there is a non-trivial interaction between the two sets of degrees of freedom, the influence functionals. Heuristically, such an effective field theory is characterized by the path integral (see \cite{Breuer:2002pc,Schlosshauer:2003zy,Sieberer:2015svu} for an overview)
\begin{equation}\label{eq:feynver}
 \int [D\Psi_{_\text{L}}]  [D\Psi_{_\text{R}}] \; \exp \bigg(i\,S_\text{s}[\Psi_{_\text{R}}] - i\, S_\text{s}[\Psi_{_\text{L}}] + i \,S_\text{IF}[\Psi_{_\text{R}}, \Psi_{_\text{L}}] \bigg) 
\end{equation}  
The quantum channel which maps density operators to density operators ought to be a completely positive trace preserving (CPTP) map on the system. In particular, the  influence functional, $S_\text{IF}[\Psi_{_\text{R}}, \Psi_{_\text{L}}] $ induced onto the system, owing to the coupling with the environment, should obey certain positivity constraints. A general proof of this statement is, as far as we are aware, not available for generic open quantum systems. The Caldeira-Leggett description of quantum Brownian motion \cite{Caldeira:1982iu} exemplifies this construction for Gaussian dynamics in quantum mechanics. Often, assuming a suitable Markovian approximation for the environment, open quantum dynamics is modeled in terms of a Linbladian, see \cite{Kulkarni:2021xsx} for an analysis in the SYK model. For finite systems, one can also obtain bounds for quantum dissipation \cite{Hayden:2021oq}. But the broad goal of constructing a local effective field theory runs into issues with perturbative interactions; see e.g., \cite{Lombardo:1995fg} for early work on the subject and \cite{Agon:2014uxa} for recent attempts in this direction, in addition to \cite{Avinash:2017asn,Agon:2017oia,Gao:2018bxz,Avinash:2019qga} for technical issues regarding renormalization. 

An important problem is \emph{to ascertain sufficient conditions for a local effective field theory to emerge}. For perturbative dynamics, this is difficult, since locality relies on the system losing memory of its interaction with the environment, 
but the relaxation timescale is long at weak coupling. Consequently, there aren't  simple microscopic models from which a local non-unitary open quantum effective  field theory has been  systematically derived.  This question can, however, be tackled directly in holographic systems, where the dynamics is intrinsically strongly coupled. Moreover, these holographic environments are fast scrambling and maximally ergodic in their dynamics, leading to a simple dynamics of probe effective field theories, as argued in \cite{Jana:2020vyx}. 

A prototype model of this is the aforementioned quantum Brownian motion, which as argued in \cite{deBoer:2008gu,Son:2009vu} can be understood as the dynamics induced on a probe quark in a thermal plasma of a  strongly coupled holographic CFT. Building on progress in real-time AdS/CFT \cite{Son:2002sd,Herzog:2002pc,Skenderis:2008dh,Skenderis:2008dg,vanRees:2009rw}, especially \cite{Glorioso:2018mmw}, there has been renewed interest and progress in this subject. Salient results to date include non-linear and non-Gaussian Langevin dynamics (including an effective field theory that probes out-of-time-order observables) \cite{Chaudhuri:2018ihk,Chakrabarty:2018dov,Chakrabarty:2019aeu}, a broad paradigm for discussing open quantum systems for probes of thermal plasmas \cite{Jana:2020vyx} and for studying systems at finite density \cite{Loganayagam:2020eue,Loganayagam:2020iol}. An interesting outcome of these explorations is a systematic construction of a  Wilsonian open effective field theory for  systems with Goldstone modes \cite{Ghosh:2020lel,He:2021jna,He:2022jnc}.

\subsection{Holographic complexity}
\label{sec:complexity2}

The future interior region of a two-sided black hole spacetime is a wormhole whose length grows with time. More specifically, the wormhole connecting the boundaries of a two-sided asymptotically AdS black hole grows linearly with boundary time at late times. The size of the wormhole can be quantified in several ways. For example, we can take the volume of a maximal bulk Cauchy slice anchored to the boundary at a given time (a codimension-1 analogue of the RT formula) \cite{Susskind:2014moa,Stanford:2014jda,Iliesiu:2021ari}, or the spacetime action of a Wheeler-de Witt patch (the causal domain of a bulk slice that reaches almost to the boundary) \cite{Brown:2015bva,Brown:2015lvg,Lehner:2016vdi}. With either definition, at late times the wormhole size grows linearly at late times with coefficient equal to the black hole's entropy.

What quantity in the boundary field theory does this growth correspond to? Susskind and collaborators proposed that it corresponds to the circuit complexity of the boundary state \cite{Susskind:2014moa,Stanford:2014jda,Brown:2015bva,Brown:2015lvg}. This idea is supported by a tensor-network picture, in which the tensor network describing the state grows with time by adding more and more legs \cite{Hartman:2013qma}. A cartoon of the time evolution of the state by a quantum circuit also shows that generically the complexity grows linearly, with coefficient given by the entropy, for times that are doubly exponential in the entropy. More quantitative evidence for the correspondence between wormhole size and complexity comes from considering shockwave geometries, in which energy is injected into the spacetime at the boundary, causing the horizon to jump outwards and the size of the wormhole therefore to increase by a discrete amount \cite{Stanford:2014jda}. On the CFT side, such injections add to the circuit necessary to prepare the state, thereby increasing its complexity.

A different set of ideas connecting holography to a notion of complexity is path-integral optimization, in which the bulk spacetime emerges from minimizing the number of operations required to prepare the state using a Euclidean path integral \cite{Caputa:2017urj,Boruch:2020wax,Boruch:2021hqs}.

While each of these proposals does seem to capture important qualitative features of holographic dualities, it remains to be seen \emph{how they are related to each other and in what regime (if any) they are correct.} The idea that circuit complexity has a simple bulk dual in holographic theories has also led to a large amount of work attempting to define and quantify this notion in general field theories, including free ones, as well as in quantum mechanics. Some of these directions were mentioned in \S\ref{sec:complexity1}.

\section{Simulation}
\label{sec:simulation}

Quantum information also plays a growing role in attempts to simulate the physics of quantum field theories. In the past, such simulations were carried out with classical computers, and quantum information has already provided a rich set of new ideas and algorithms for classical simulation. Moreover, we now have the prospect of simulating quantum field theories and quantum gravity with quantum computers. We discuss these two cases in turn, starting with classical simulation.

\subsection{Classical simulation}

Take the case of Monte Carlo methods, a widespread and powerful suite of tools for classical simulation of field theories. These methods allow efficient non-perturbative access to Euclidean path integrals when the path integral does not possess a sign problem. Yet sign problems are common, especially in systems with fermions, in systems at finite density, and in the context of real-time dynamics (e.g.~\cite{nagata2021finitedensity}). Given the serious constraints on Monte Carlo methods imposed by the sign problem, it is crucial to ask whether the sign problem has larger physical significance. For example, is the sign problem a signal that a system necessarily requires a quantum simulation to efficiently deal with it? Remarkably, the answer is no: there is no conservation of difficulty and alternate methods that leverage structure in the physics can completely bypass the sign problem. In particular, tensor network methods for classical simulation provide a key example where entanglement ideas have led to powerful new classical simulation tools.

Tensor networks rose to prominence in the many-body physics community, especially in the context of systems with one spatial dimension. The simplest kinds of tensor networks in widespread use are 1d matrix product states (MPS), and MPS methods have now evolved to the point where they almost provide black-box routines for 1d physics. This includes everything from low-temperature equilibrium physics to high-temperature charge and energy transport, and the key is an in-depth understanding of the entanglement structure of these systems. Given the many similarities between many-body models and lattice regulated field theories (in the Hamiltonian formalism), it is quite plausible that these methods will have a major impact on field theory problems as well. Promising early steps have already been taken, and there remains much to understand (see \cite{banuls_2020} for a review).

One class of phenomena where tensor networks have been surprisingly effective is the description of transport phenomena in strongly-coupled systems. For these problems, we typically know that the IR effective theory should be hydrodynamics, but matching to the UV is difficult. This is because computing the transport coefficients requires understanding the microscopic quantum dynamics out to hydrodynamic timescales. Remarkably, by adapting the setup to reduce entanglement generation, either with specially designed coarse-grainings of the dynamics~\cite{White_2018,rakovszky2020dissipationassisted,white2021effective} or by using open systems~\cite{Prosen_2009}, tensor networks have given new access to transport properties of strongly interacting 1d systems. A major open question is \emph{how to extend these ideas to low temperatures~\cite{Zanoci_2021} and to 2d and beyond.}

\subsection{Quantum simulation}

These developments in classical simulation are changing the landscape of what is possible, but it still seems inevitable that quantum simulation will also be called for. Indeed, many physical processes presumably require an inherently quantum computational approach to simulation. And even for problems where we have a plausible route to classical simulation, there is usually significant overhead associated with using a classical description, so that for systems in higher dimensions and with more degrees of freedom, quantum simulation will be necessary.

When discussing quantum simulation, a standard framework is to think about so-called quench experiments where one prepares an initial state, evolves it in time, and then makes a final measurement. Of these components, initial state preparation is often the most challenging~\cite{Jordan_2012}. One reason is that, while we are familiar with preparing field theory states using Euclidean path integrals, the associated imaginary time dynamics is not natural on a quantum device.

For example, if one is interested in scattering, then the initial state could correspond to a ground state perturbed by a small number of excitations \cite{Jordan_2012,2020arXiv201207243M,2021PhRvL.127u2001B}. Alternatively, Gibbs states can be used to describe processes at non-zero temperature and density, such as might be relevant in the early universe or in heavy-ion collisions. The notions of state complexity discussed in \S\ref{sec:complexity1} and \S\ref{sec:complexity2} are particularly relevant here, since more complex states by definition require more resources to prepare. It is important to find methods to prepare field theory states and to estimate the resources required. Here again, tensor networks provide an interesting technology, for example, MERA and DMERA tensor networks have been shown to be able to represent simple UV completions of field theories and can be adapted to give explicit quantum circuit prescriptions for building these states, e.g.~\cite{2011arXiv1109.5334E,2021ScPP...10..143W,2017arXiv171107500K,2021arXiv210909787S,2017PhRvL.119a0603H}. \textit{However, much remains to be understood here, especially when dealing with gauge symmetry and related ideas, e.g.~\cite{2021PhRvD.104g4505D,2020arXiv201106576B,2020PhRvD.102i4501M}.}

In the context of resource estimates, the formal framework of resource theories is useful. Entanglement is one well-known example of a resource, but there are many others. One that has received attention very recently is known as magic 
\cite{BravyiKitaev2005}. This is a resource for qubits analogous to non-Gaussianity in field theory, with non-magical states corresponding to a special class known as stabilizer states (very important in error correction) and with magic being a potentially relevant resource for some fault tolerant quantum computation schemes. Preliminary studies of magic in many-body systems and lattice-regulated field theories have been undertaken, with one result being that a simple 2d CFT was shown to be highly magical~\cite{White:2020zoz}. \textit{Here too, much remains to be explored, especially in formulating more continuum-friendly notions of magic and other resources and in designing new simulation methods, classical and quantum, that take advantage of the structure of magic in states of interest.} These methods should also connect to older ideas exploring the phase space representation of QFT and the use of Wigner functions, e.g.~\cite{Mrowczynski_1994}, and more recent studies of negativity, e.g.~\cite{2021arXiv211010736K}.

Looking broadly, all of these approaches to simulation stand to teach us a great deal about the physics of strongly coupled field theories and quantum gravity. The value of simply understanding how in principle to formulate such problems as tractable simulations should not be underestimated, similar to the way Wilson's deep insights into renormalization were catalyzed by asking how one could at least in principle simulate QFT on a classical computer. In terms of particular applications, strongly coupled dynamics is one area where simulations will plausibly have a major impact, especially as quantum and hybrid classical-quantum methods give access to more complex systems. Another area where we expect significant impact is simulations of quantum gravity. For example, holographic duality predicts surprising phenomena from the point of view of QFT and quantum information, such as teleportation via traversable wormholes, that might teach us more about the requirements for spacetime to emerge from microphysics~\cite{2017JHEP...12..151G,2017ForPh..6500034M,2018arXiv180400491M,2019arXiv191106314B,2021arXiv210201064N}. One can also investigate similar holographic simulations of string theory, e.g.~\cite{2021JHEP...07..140G}. 

We should also mention the very real prospect that better understanding of field theories may feed back into quantum information science. For example, the toric code, which is a leading candidate for error correction in various concrete fault-tolerant schemes, is in essence a $Z_2$ gauge theory in the extreme deconfined limit~\cite{Kitaev_2003}. \textit{Other kinds of strongly coupled gauge theories, especially those of relevance for quantum gravity~\cite{Almheiri:2014lwa}, might bring additional insights and models for quantum error correction.}

\section{Outlook}
\label{sec:outlook}

As we have described throughout this white paper, there has been remarkable progress in the past decade or so in our understanding of quantum information in QFTs. Nonetheless, many important questions remain to be addressed. Some of these we have noted in the context of our discussion in the text. Here we outline some of the big-picture questions, where we hope progress will be made over the next few years.

\paragraph{What is QFT?:} Despite its wide success as a framework for understanding the fundamental principles of nature, it is fair to say that we don't yet fully understand the nature of QFT. 
In addition to the growing role of information-theoretic methods applied to QFT reviewed in this white paper, the past decade has also seen new geometric approaches to computing observables such as  scattering amplitudes, further development of  non-perturbative bootstrap techniques, a better understanding of symmetries and charges,  and ever deeper connections to many areas of mathematics. All of these developments have in different ways lent insights far transcending the standard textbook treatments of QFT. One natural question is: Which of these ingredients is a defining feature of the theory? What is the canonical toolkit of  a future quantum field theorist, as presented in a textbook a couple of decades hence? While these questions might be intangible at present, part of the progress will likely come from asking questions about the interconnections among these developments. Some of these are easier to fathom, being already somewhat developed, such as the connection between symmetry charges and operator algebraic formulations of QFT. On the other hand, while we have several different proofs of quantities that are RG monotones ($c$, $F$, and $a$ theorems), as yet we still lack a unified picture.  Nevertheless, understanding how the information theoretic data in QFTs relates to other observables, and developing further techniques to extract them, could provide valuable insight towards this goal.

\paragraph{Gauge invariance and information:}  In theories with gauge-invariant degrees of freedom, the decomposition of algebras comes with non-trivial centers and related superselection sectors. Several open questions remain in this context. What is the significance of algebraic centers for lattice gauge theories in the continuum limit?  
What is the role of topological entanglement in gapless theories? How do we describe gauge theory entanglement in the algebraic approach? What of interacting theories in the algebraic approach? Does the change in the nature of the von Neumann algebras, noted in the case of large $N$ confinement-deconfinement transition \cite{Leutheusser:2021frk,Witten:2021unn}, lead to new insights into the dynamics of confining gauge theories?

\paragraph{From fields to gravity and strings:} In the gravitational context, we now understand, at the semiclassical level, the relevance of the generalized entropy, an object that combines a Bekenstein-Hawking-like classical term with a quantum von Neumann entropy. It has played a crucial role both in the development of gravitational entropy bounds, and in the recent discussions of the black hole information paradox.  An open problem is to better understand this quantity beyond the semiclassical regime. How does one pick subregions and define their algebras in this setting in a relational manner, 
maintaining diffeomorphism invariance? How does one  define these quantities in string theory? 
Can the information theory approach shed light on the question of which effective field theories can be consistently completed into a quantum theory of gravity, and which lie in the swampland? In the context of holography, it is interesting to ask about the information-theoretic nature of QFT wavefunctions. In particular, which aspects of their entanglement structure are central for the holographic entropy inequalities, and what does this tell us about quantum gravity?  Is there a principled way to discern which of the various conjectures relating to complexity are valid, and what  they are telling us about the nature of quantum gravitational wavefunctions?

\paragraph{Real-time dynamics and cosmology:}  QFT dynamics in cosmological spacetimes shares many characteristics of open quantum systems. As outlined above, to date, there isn't a clean description of the effective dynamics of quantum fields in an open system, and many questions remain to be addressed. Progress in these directions should help us better formulate the issues we need to confront in the cosmological context, be it the nature of observables, the temporal evolution of information-theoretic quantities such as von Neumann entropy, and so on. It would furthermore be interesting to tie these to the study of cosmological correlators, which has been tackled using bootstrap methods.

\ \\

\ \\

\noindent \textbf{Acknowledgments}\\
We thank V.\ Hubeny and L.\ Rastelli for encouraging and guiding the writing of this white paper. We also thank H.~Casini and T.~Takayanagi for useful comments on a draft. TF is supported by the Air
Force Office of Scientific Research under award number FA9550-19-1-036
and by the DOE award number DE-SC0019183. TH is supported by the Simons Foundation and NSF grant PHY-2014071. MH is supported by the Simons Foundation through the It from Qubit Collaboration and by the DOE through grant DE-SC0009986.
MR is supported by  DOE grant DE-SC0009999 and by funds from the University of California. MH and MR are also supported by DOE grant DE-SC0020360 under the HEP-QIS QuantISED program. The work of BS is supported in part by the AFOSR under grant  FA9550-19-1-0360.

\bibliographystyle{jhep}
\bibliography{ref}
\end{document}